\documentclass[pdflatex,sn-mathphys-num]{sn-jnl}% Math and Physical Sciences Numbered Reference Style
\usepackage{booktabs}  % for top/mid/bottom rules
\usepackage{float}  
\usepackage{graphicx}%
\usepackage{multirow}%
\usepackage{amsmath,amssymb,amsfonts}%
\usepackage{amsthm}%
\usepackage{mathrsfs}%
\usepackage[title]{appendix}%
\usepackage{xcolor}%
\usepackage{textcomp}%
\usepackage{manyfoot}%
\usepackage{booktabs}%
\usepackage{algorithm}%
\usepackage{algorithmicx}%
\usepackage{algpseudocode}%
\usepackage{listings}%
\theoremstyle{thmstyleone}%
\theoremstyle{thmstyletwo}%

\theoremstyle{thmstylethree}%

\begin{document}

\title[Realization of Graphene Quantum Dots for Innovative Biosensor Development and Diverse Applications] {Realization of Graphene Quantum Dots for Innovative Biosensor Development and Diverse Applications}


\author*[1]{\fnm{Kumar Gautam}}\email{edumonk@ieee.org}

\author[2]{\fnm{Kumar Shubham}} 

\author[3]{\fnm{Hitesh Sharma}} 

\author[4]{\fnm{Divya Punia}} 

\author[5]{\fnm{Ajay K Sharma}} 

\author[6]{\fnm{Namisha Gupta}} 

\author[7]{\fnm{Varun Rathor}} 

\author[8]{\fnm{Vishakha Singh}}

\affil[1, 2, 7]{\orgdiv{Department of Quantum System and Sensing}, 
\orgname{Egreen Quanta LLP, Delhi\\Quantum Research And Centre of Excellence}, \city{New Delhi}, \postcode{110036}, \state{Delhi}, \country{India}}

\affil[3,4,5,6,8]{\orgdiv{Department of Electronics and Communication Engineering}, 
\orgname{National Institute of Technology Delhi}, 
\orgaddress{\street{Institute Campus}, \city{New Delhi}, \postcode{110036}, \state{Delhi}, \country{India}}}


\abstract{This paper investigates quantum dots (QDs), which are miniature semiconductor structures with remarkable optical and electrical properties due to quantum confinement processes. Traditional QDs, such as CdTe, have been extensively investigated; however, they frequently exhibit toxicity and stability issues. Graphene quantum dots (GQDs) are emerging as a safer and more stable alternative to traditional QDs. GQDs are honeycomb-lattice carbon atoms with unique electronic and optical properties that make them promising candidates for biomedical, electronic, and energy storage applications. GQD synthesis methods (top-down and bottom-up) and their advantages over standard QDs include better photostability, biocompatibility, and configurable band gaps. GQDs are perfect for real-world uses like sensitive biosensing, real-time food safety monitoring, and smart packaging because of their low toxicity, high sensitivity, and affordability. These uses are all essential for cutting down on food grain waste. This emphasizes the growing significance of GQDs in advancing nanotechnology and their potential integration with quantum technologies, paving the door for creative solutions in biosensing, food safety, environmental monitoring, and future quantum electronics.}




\maketitle

\section{Introduction}\label{sec1}
Quantum dots (QDs) are zero-dimensional semiconductor nanocrystals typically ranging from 2--10 nm in diameter, containing 100--10,000 atoms \cite{parak2010fundamental}. At the nanoscale, these particles demonstrate distinct electronic and optical behaviors resulting from quantum confinement, which limits electron movement in all three dimensions. QDs are particularly notable for bridging the properties of bulk materials and single atoms or molecules, enabling their characteristics to be tailored through modifications in size, composition, or surface functionalization. They are novel nanocrystalline semiconductor materials whose electrical as well as optical properties strongly depend upon the size and shape of the dots. Due to their very large surface-to-volume ratio, QDs possess characteristics intermediate between individual atoms/molecules and bulk materials. QDs can be produced either from a single element (e.g., silicon, germanium) or from composite elements (such as CdSe, CdS, and so on) \cite{kim201325th}. Their size-dependent emission further enhances their utility: smaller QDs emit blue light (460 nm), whereas larger ones emit red light (650 nm) \cite{kim201325th,wang2013energy}.
In recent years, the novelty of Graphene Quantum Dots (GQDs) has gained significant attention, positioning them as a next-generation alternative to conventional semiconductor-based QDs in biosensing. Unlike heavy-metal-based QDs (e.g., CdTe, CdSe), GQDs exhibit low toxicity and excellent biocompatibility, making them suitable for direct biomedical and food-related applications. Their tunable bandgap enables versatile detection across a wide range of biological and environmental targets, while their superior photostability ensures reproducibility and long-term assay stability. Furthermore, their cost-effective and eco-friendly synthesis facilitates scalable production and sustainable deployment. Importantly, GQDs exhibit ultra-high sensitivity, detecting trace levels of pathogens and toxins, and their seamless scalability allows integration into portable biosensors, Internet of Things (IoT) systems, and even quantum-enabled sensing platforms. Collectively, these attributes underline the novel role of GQDs as transformative materials for next-generation biosensing technologies.\\Several studies have demonstrated the remarkable versatility of GQDs in optical and electrochemical biosensing. For instance, their strong photoluminescence and excitation-wavelength-dependent emission facilitate label-free detection of biomolecules such as DNA, proteins, and glucose \cite{liu2015graphene}. Functionalization of GQDs with specific biomolecules or recognition elements significantly enhances selectivity and sensitivity, allowing real-time monitoring in complex biological matrices. Moreover, the edge chemistry of GQD rich in carboxyl, hydroxyl, and amino groups—provides a robust platform for covalent attachment of biomolecules, thereby improving sensor performance and stability \cite{rasheed2024graphene}.

In addition to biosensing, GQDs have shown potential in theranostic applications, merging diagnostic and therapeutic functionalities. Their photothermal and photodynamic properties allow simultaneous imaging and targeted therapy, highlighting their multifunctional capabilities beyond conventional QDs \cite{liu2015graphene}. Recent advancements also emphasize the development of hybrid GQD composites, combining graphene derivatives with metal nanoparticles or polymers, which further enhance signal amplification and broaden the detection range for environmental and clinical monitoring \cite{rasheed2024graphene}. These developments underscore the growing trend of integrating GQDs into multifunctional sensing platforms that combine high sensitivity, selectivity, and portability.

\subsection{Biosensing Applications of Graphene Quantum Dots}

Beyond their physicochemical advantages, GQDs have demonstrated remarkable application potential in quantum biosensing across medical, food safety, environmental, and IoT domains. In medical diagnostics, QD-magnetic nanoparticle hybrids can isolate and detect circulating tumor cells with sensitivity down to 21 cells/mL, enabling detection of lung cancer micro-metastases, while multi-color QDs facilitate multiplex biomarker screening at femtomolar levels \cite{iannazzo2021smart}. For cardiovascular monitoring, GQD-based electrochemical sensors have achieved troponin detection limits as low as 0.1 pg/mL in serum, with rapid assays completed in under 10 minutes performance comparable to laboratory-based methods and highly relevant for point-of-care diagnostics \cite{tabish2022graphene}.  

In the field of food safety, GQD fluorescent aptasensors have achieved exceptional performance for mycotoxin detection. For aflatoxin B1 (regulated at 15 $\mu$g/kg by FSSAI and 4-20 ng/mL by EU standards), GQD-based immunosensors achieve detection limits of 0.03 ng/mL in standard samples and 0.05 ng/g in contaminated maize \cite{bhardwaj2018graphene}, values that are more than two orders of magnitude lower than regulatory thresholds. Electrochemical aptasensors further demonstrate sensitivity of 213.88 $\Omega$ (ng mL$^{-1}$)$^{-1}$ cm$^{-2}$, excellent reproducibility with RSD $<$5\%, and recovery rates ranging from 80.2-98.3\% in real food matrices \cite{chen2022recent}. GQD/AuNP nanocomposites have pushed detection limits down to 0.008 ng/mL for aflatoxin B1 in maize, while maintaining 94\% current signal retention even after four weeks of storage at 4$^{\circ}$C \cite{chen2022recent}. For aflatoxin M1, MoS$_2$ QDs/UiO-66-modified sensors detect as low as 0.06 ng/mL across a linear range of 0.2-10 ng/mL \cite{kaur2022selective}. Such enhanced performance is enabled by GQDs’ fast electron transfer kinetics ($97.63 \times 10^{-5}$ cm s$^{-1}$), rapid response times (0.005 s), and strong selectivity, with recovery rates above 90\% in maize, milk, and wheat \cite{bhardwaj2018graphene,kaur2022selective,chen2022recent}. Aptamer--GQD hybrids also enable zearalenone detection with 95\% selectivity \cite{anh2019sulfur}. Importantly, this high performance is achieved with material costs of only \$10-50/g for GQDs compared to \$100-500/g for CdTe QDs, highlighting both sensitivity and cost-efficiency.  

In environmental monitoring, GQDs demonstrate the ability to detect heavy metals such as Pb$^{2+}$ at concentrations as low as 0.6 nM with response times of just 3 seconds \cite{chung2021graphene}, as well as nanomolar-level detection of organophosphate pesticides. Their integration into IoT platforms further extends these capabilities: GQD/polyimide composite humidity sensors show a 96.36\% improvement in sensitivity compared to conventional sensors with response times of 15 s, while GQD-ZnO nanocomposites detect CO across 1-100 ppm with 90\% accuracy \cite{hou2022advanced,griggs2016graphene}. When combined with low-power long-range communication technologies such as LoRaWAN, these sensors can be deployed in grain warehouses with centralized dashboards that issue real-time alerts when environmental thresholds (e.g., humidity $>$70\%, temperature $>$25$^{\circ}$C) are exceeded, thereby enabling automated control systems for ventilation or drying.  

Looking forward, GQDs also play a central role in the emerging field of quantum-optimized sensor networks. By employing algorithms such as Quantum Particle Swarm Optimization (QPSO), sensor placement can be optimized in large-scale facilities for example, reducing the number of nodes from 500 to 150 in a 10,000 m$^2$ warehouse, corresponding to a 70\% cost reduction \cite{chung2021graphene}. Moreover, QPSO reduces data transmission energy consumption by 40\% compared to classical methods, while the Quantum Approximate Optimization Algorithm (QAOA) prioritizes data routing from high-risk zones, such as warehouse corners prone to moisture migration \cite{ratre2022bioanalytical}. These advances highlight not only the adaptability of GQDs for current biosensing needs but also their potential as key enabling materials for future quantum-enhanced, IoT-integrated sensing ecosystems.  

\subsection{Quantum Confinement}
Quantum confinement refers to the change in a particle's properties when it is limited to an extremely small region. This effect, noticeable in nanoscale systems, restricts particle movement along one or more dimensions, resulting in altered electronic properties not seen in larger structures and leading to discrete, quantized energy states. Due to the quantum confinement phenomenon in quantum dots, the motion of charge carriers (electrons and holes) is restricted in all three spatial dimensions\cite{parak2010fundamental}. This confinement leads to discrete energy levels like atoms and results in size-dependent optical and electronic behaviors, such as tunable fluorescence and bandgap energies\cite{parak2010fundamental}. Some keynotes regarding confinement:
\begin{enumerate}[I.]
    \item In the case of \textbf{strong confinement}, the radius of quantum dots is less than the Bohr radius of the exciton.
    \item In the case of \textbf{moderate confinement}, the radius of quantum dots is less than the Bohr radius of either the electron or the hole.
\end{enumerate}
The extent of the confinement also affects the spatial overlap of the electron and hole wavefunctions and the Coulombic interaction between them. The Brus equation describes the relationship between QD size and bandgap energy:
\begin{equation}
\Delta E(r) = E_{\text{gap}} + \frac{h^{2}}{8r^{2}} \left( \frac{1}{m_{e}} + \frac{1}{m_{h}} \right),
\end{equation}
where $r$ is the QD radius, and $m_{e}$ and $m_{h}$ are the effective masses of electrons and holes, respectively\cite{wang2013energy}.  

\begin{figure}[h!]
    \centering
    \includegraphics[width=0.8\textwidth]{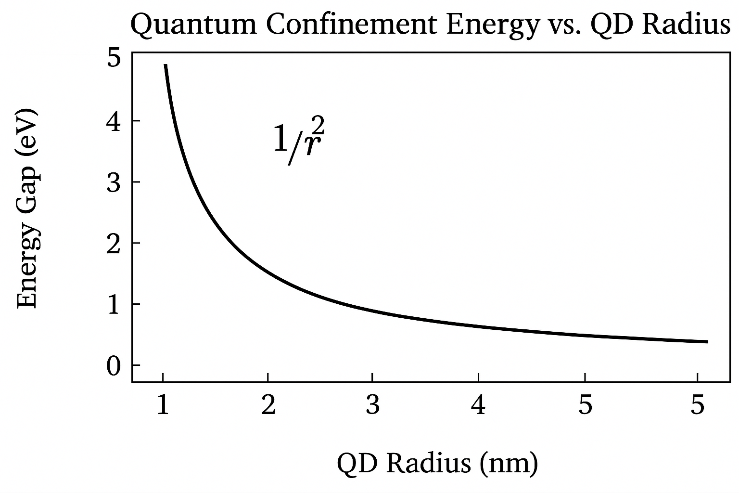}
    \caption{I-V Characteristics Showing Coulomb Blockade Staircase}
    \label{fig:qd_coulomb}
\end{figure}

As shown in Figure \ref{fig:qd_coulomb}, this graph illustrates the principle of quantum confinement in quantum dots (QDs). It demonstrates that the energy gap is inversely proportional to the square of the QD radius ($E_{g} \propto 1/r^{2}$). As the quantum dot becomes smaller, its energy gap increases, which allows for tuning its electronic and optical properties simply by changing its size.

\subsection{Surface-to-Volume Ratio}

\noindent \textbf{Mathematical Relationship:} \\
For spherical QDs, the surface-to-volume ratio scales as,
\begin{align}\frac{S}{V} = \frac{6}{D}\end{align}
where $D$ is the diameter. As diameter decreases from $1 \, \text{m}$ to $10 \, \text{nm}$, 
$\tfrac{S}{V}$ increases by 100-fold, from $6 \times 10 \, \text{m}^{-1}$ to $6 \times 10^{8} \, \text{m}^{-1}$\cite{bera2010quantum}. \\[6pt]
\noindent \textbf{Impact on Biosensing Properties:}
\begin{itemize}
    \item \textbf{Enhanced binding capacity:} $\sim 15\%$ of atoms are surface exposed in 5 nm CdS QDs, providing abundant sites for bioconjugation\cite{bera2010quantum}.
    \item \textbf{Increased reactivity:} High surface atom density enables multiple probe attachments per QD.
    \item \textbf{Surface trap states:} Can affect optical properties; passivation with ZnS shells improves quantum yields.
\end{itemize}

\subsection{Voltage/Current Characteristics}

\textbf{Coulomb Blockade Effects}
Quantum dots (QDs) exhibit single-electron transistor behavior due to electrostatic repulsion. Key characteristics include:
\begin{itemize}
    \item \textbf{Staircase I-V curves:} Current increases in discrete steps as bias voltage increases.
    \item \textbf{Charging energy:} Typically 1--10~meV in semiconductor QDs, given by
    \[
        E_C = \frac{e^2}{2C}
    \]
    where $C$ is the dot capacitance~\cite{mcardle2024new}.
    \item \textbf{Diamond-shaped stability regions:} Seen in differential conductance maps, indicating electron number quantization.
\end{itemize}

\textbf{Practical Measurements}
\begin{itemize}
    \item \textbf{CdTe QD diodes:} Show linear I--V characteristics at room temperature with on/off ratios of approximately 200:1~\cite{gumaste2021iv}.
    \item \textbf{PbS QD chemiresistors:} Exhibit carrier mobility of $\sim 0.07~\text{cm}^2/\text{V}\cdot \text{s}$ and gas sensitivity ratios of 15:1 for 30~ppm NO$_2$~\cite{gumaste2021iv}.
    \item \textbf{Temperature dependence:} Coulomb blockade effects are most prominent below 0.1~K.
\end{itemize}

\section{Graphene-Based Quantum Dots (GQDs)}
Graphene, consisting of interconnected sp² carbon atoms, offers an array of remarkable features such as excellent electrical conductivity, mechanical robustness, extensive surface area, high thermal stability, and flexibility. From the growing class of quasi-zero-dimensional graphene-based nanomaterials, GQDs have attracted considerable research interest.

GQDs exhibit unique size-dependent properties that reflect the combined structural characteristics of small graphene sheets, enhanced with quantum confinement and edge effects~\cite{parak2010fundamental}. Compared to bulk graphene, GQDs offer several additional advantages, including tunable bandgap, excellent aqueous solubility, good biocompatibility, high fluorescence quantum yield, and the presence of multiple active functional groups offering unprecedented opportunities in the field of nanodevices~\cite{mamba2020graphene}.

GQDs are considered tiny fragments of graphene and are zero-dimensional derivatives of bulk carbon materials, combining properties of both carbon dots and graphene~\cite{mamba2020graphene}. An ideal GQD consists of a single atomic layer of carbon atoms, although the lateral dimensions may vary. In practice, however, most synthesized GQDs have multiple atomic layers with lateral sizes typically less than 10~nm and frequently contain surface functional groups such as oxygen and hydrogen~\cite{parak2010fundamental}.

\begin{figure}[h!]
    \centering
    \includegraphics[width=0.8\textwidth]{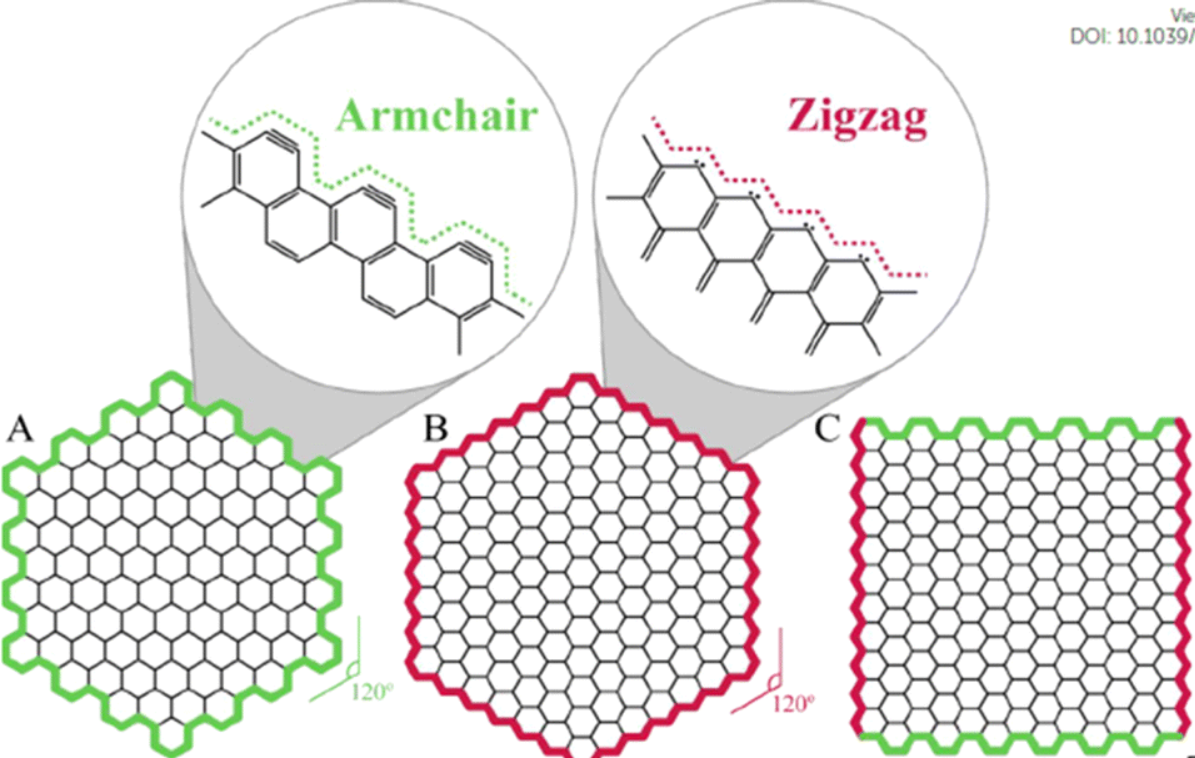} 
    \caption{The shapes of graphene quantum dots depend on the type of edge: (A) armchair, (B) zigzag, and (C) hybrid armchair-zigzag GQDs. The amplified images in (A) and (B) show structural details of the armchair and zigzag edges, respectively. Reproduced with permission from Ref.82. Copyright © Elsevier 2020.}
    \label{fig:2}
\end{figure}

In GQDs, graphene is confined to nanoscale dimensions, and quantum confinement effects give rise to a tunable band gap. The band gap of GQDs depends on several factors, the most significant of which is their physical size (see Fig.~\ref{fig:2}). This size-dependent quantum confinement results in unique opto-electronic properties.

The structural characteristics of GQDs enable them to exhibit remarkable properties, including low toxicity, stable photoluminescence, high chemical stability, and strong quantum confinement effects~\cite{mamba2020graphene}. Their fluorescence emission spans a broad spectral range from ultraviolet through visible to infrared. The precise origin of this photoluminescence remains an active area of research, with mechanisms proposed such as quantum confinement, surface defect states, and pH-dependent functional groups~\cite{mamba2020graphene}.

Furthermore, the electronic structure of GQDs is critically influenced by the crystallographic orientation of their edges. For example, zigzag-edged GQDs with diameters of 7--8~nm have been observed to exhibit metallic behavior. Generally, their energy gap decreases with an increase in the number of graphene layers or carbon atoms per layer (see Fig.~\ref{fig:2})~\cite{mamba2020graphene}.

\subsection{Why Did We Choose Graphene?}
There are several unique properties of graphene that make it particularly suitable for QD fabrication:

\begin{itemize}
    \item \textbf{Large Surface Area:} A theoretical surface area of 2630~m$^2$/g allows for increased surface interactions in sensor and catalysis applications~\cite{liu2015graphene}.
    \item \textbf{High Electron Mobility:} Electron mobility values ranging from 10,000 to 15,000~cm$^2$/Vs enable ultrafast charge transport, which is advantageous for electronic and optoelectronic applications~\cite{chinnusamy2018incorporation}.
    \item \textbf{Atomic Thickness:} Graphene has an atomic thickness of approximately 0.34~nm and offers a tunable energy gap ranging from 0.5 to 7~eV~\cite{liu2015graphene}.
    \item \textbf{Biocompatibility and Low Toxicity:} Graphene and its derivatives exhibit excellent compatibility with biological systems, making them viable for biomedical applications.
    \item \textbf{Edge-Defect-Driven Photoluminescence:} Enables multicolor emission through edge states and structural defects.
    \item \textbf{Size and Edge Dependence:} GQD properties depend significantly on their size (typically 2--20~nm) and edge structure, with zigzag and armchair terminations affecting electronic behavior~\cite{liu2015graphene}.
    \item \textbf{Stable Photoluminescence (PL):} Exhibits high PL quantum yields—up to 94\% and strong resistance to photobleaching~\cite{mamba2020graphene}.
    \item \textbf{Functional Group Availability:} Oxygen-rich groups (e.g., --COOH, --OH) facilitate covalent bonding with biomolecules, enhancing biosensing capabilities.
\end{itemize}

\subsection{Properties of GQDs}
\begin{figure}[h!]
    \centering
    \includegraphics[width=0.8\textwidth]{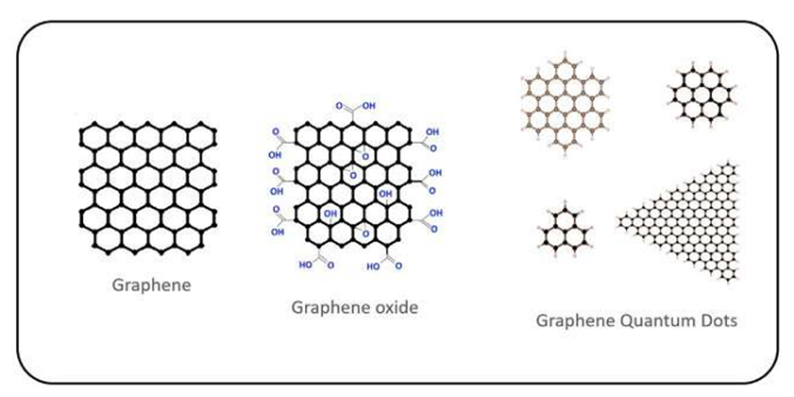} 
    \caption{Schematic illustration of the structure of (a) Graphene, (b) Graphene Oxide, and (c) Graphene Quantum Dots, which are nanoscale fragments of a graphene sheet.}
    \label{fig:3}
\end{figure}

As shown in Fig.~\ref{fig:3}, GQDs are distinct from their parent materials, pristine graphene and graphene oxide (GO), which is a larger sheet functionalized with oxygen-containing groups. In GQDs, carbon atoms are organized in a hexagonal honeycomb lattice. Each carbon forms three covalent bonds, with $\pi$ electrons delocalized perpendicular to the plane. The edges can exhibit diverse bonding patterns, including carbyne-like terminal triple bonds or carbene-like configurations with lone electron pairs at the edges.

These edge configurations significantly influence the overall shape of the GQDs. As illustrated in Fig.~\ref{fig:qd_coulomb}, when identical edge types meet, a 120$^\circ$ angle is formed, resulting in a hexagonal geometry. Conversely, when differing edge types meet, a 90$^\circ$ angle is formed, producing rectangular GQDs~\cite{mamba2020graphene,balkanloo2023graphene,mao2013manipulating}. Edge configuration is thus a critical structural feature that governs the electronic and optical properties of GQDs.

GQDs typically show absorption bands in the deep ultraviolet (UV) region and often extend into the UV-visible spectrum. Their absorption and luminescence characteristics can vary with surface morphology and chemical modifications. Unlike traditional semiconductor QDs that require 40--90\% doping for high quantum yields, GQDs can achieve quantum yields as high as 94\%~\cite{chung2021graphene}.

The photoluminescence (PL) characteristics of GQDs are influenced by both $\pi \to \pi^*$ transitions and $n \to \pi^*$ transitions, with the latter typically resulting in lower-energy, longer-wavelength emissions. By controlling the synthesis temperature, GQDs were produced in three distinct size ranges: 1--4~nm, 4--8~nm, and 7--11~nm~\cite{chung2021graphene}.

As the size of the GQDs increased, the energy bandgap gradually decreased, resulting in a photoluminescence shift from blue to green to yellow~\cite{chung2021graphene}. Single-layer GQDs, with an average height of approximately 0.5~nm, exhibited yellow photoluminescence, whereas multilayered GQDs (2--6 layers, 1--3~nm height) appeared green~\cite{chung2021graphene}. The bandgap energy decreases as the GQD size increases.

Furthermore, GQDs with zigzag edge conformations tend to have smaller bandgaps than those with armchair edges. The synergy of high conductivity, large surface area, and chemically active edge sites enables GQDs to exhibit excellent electrochemical activity~\cite{chung2021graphene}.

\subsection{Synthesis Methods}
\begin{figure}[h!]
    \centering
    \includegraphics[width=0.8\textwidth]{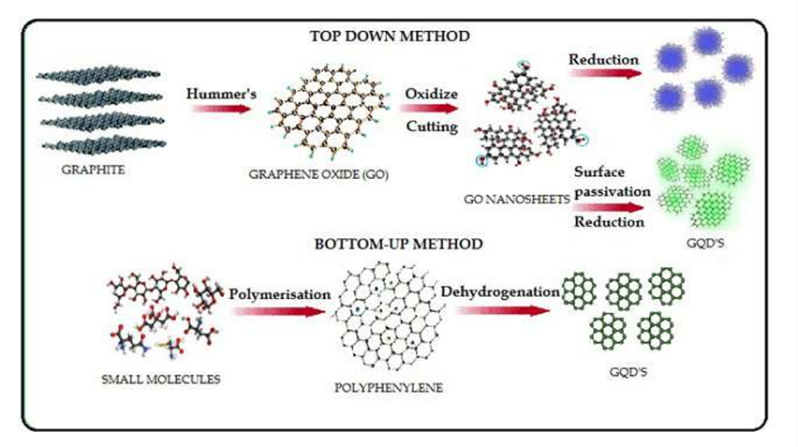} 
    \caption{Schematic illustration of top-down and bottom-up approaches for synthesis of GQDs.}
    \label{fig:4}
\end{figure}

The synthesis of GQDs can be broadly categorized into two main approaches, as illustrated in Fig.~\ref{fig:4}. The \textit{top-down} method involves breaking down larger carbon materials like graphite or graphene oxide into smaller fragments. Conversely, the \textit{bottom-up} method builds GQDs from the ground up by polymerizing small molecular precursors into a larger nanostructure.

\subsubsection*{Top-Down Methods}
\begin{enumerate}
    \item \textbf{Chemical Oxidation}
    \begin{itemize}
        \item \textit{Process:} Bulk carbon materials (e.g., graphite, carbon nanotubes) are oxidized using strong acids (e.g., H$_2$SO$_4$/HNO$_3$) to fragment into GQDs.
        \item \textit{Output:} GQDs of 1--10~nm with oxygen-containing functional groups (--COOH, --OH)~\cite{roushani2015graphene}.
        \item \textit{Drawback:} Generates toxic waste.
    \end{itemize}
    
    \item \textbf{Hydrothermal/Solvothermal}
    \begin{itemize}
        \item \textit{Process:} Carbon precursors (e.g., biomass, citric acid) are heated at 150--200$^\circ$C under high pressure.
        \item \textit{Example:} Tian et al. synthesized GQDs (1--1.5~nm thick, 20--40~nm diameter) with 15\% quantum yield (QY) using H$_2$O and DMF~\cite{rasheed2024graphene}.
        \item \textit{Yield:} Preparation yields range from 9--13\%, and functionalization (e.g., with PEG) increases yield to $\sim$28\%~\cite{rasheed2024graphene}.
        \item \textit{Advantages:} Simple, scalable, and eco-friendly.
    \end{itemize}
    
    \item \textbf{Electrochemical Exfoliation}
    \begin{itemize}
        \item \textit{Process:} Graphite electrodes are exfoliated in electrolytes (e.g., citric acid + NaOH) under an applied voltage.
        \item \textit{Output:} Size-controlled GQDs (2--3~nm) with high photoluminescence QY (82\%)~\cite{rasheed2024graphene}.
        \item \textit{Advantage:} Green method with no toxic reagents.
    \end{itemize}
    
    \item \textbf{Pulsed Laser Ablation (PLA)}
    \begin{itemize}
        \item \textit{Process:} Laser irradiation fragments graphite flakes in solvents (e.g., chlorobenzene).
        \item \textit{Output:} Pristine GQDs (1.4--4.2~nm) with minimal defects~\cite{kaushal2022short}.
    \end{itemize}
\end{enumerate}

\subsubsection*{Bottom-Up Methods}
\begin{enumerate}
    \item \textbf{Microwave-Assisted Synthesis}
    \begin{itemize}
        \item \textit{Process:} Microwave radiation rapidly carbonizes organic precursors (e.g., glucose) in minutes.
        \item \textit{Output:} Uniform GQDs (3--5~nm) with tunable emission properties~\cite{kaushal2022short}.
    \end{itemize}
    
    \item \textbf{Organic Precursor Carbonization}
    \begin{itemize}
        \item \textit{Process:} Small molecules (e.g., citric acid) are carbonized at high temperatures.
        \item \textit{Advantage:} Enables precise control over size and heteroatom doping (e.g., nitrogen-doped GQDs).
        \item \textit{Drawback:} Requires complex optimization procedures.
    \end{itemize}
\end{enumerate}

\section{Mathematical Model of GQDs}

GQDs are finite-sized graphene fragments where quantum confinement effects dominate, leading to discrete energy levels and a tunable bandgap unlike bulk graphene's zero-gap semimetal behavior. A common mathematical model for GQDs uses the low-energy effective Dirac equation, treating electrons as massless Dirac fermions. This model is particularly useful for circular GQDs with infinite mass confinement (a common approximation for hard-wall boundaries). Below, we derive the key equations step by step, focusing on the electronic structure and energy levels in the absence of external fields for simplicity. (For magnetic fields or flux, extensions exist, but we note them briefly.)

\subsection{ Dirac Hamiltonian for Graphene}

The starting point is the low-energy approximation near the Dirac points ($K$ and $K'$ valleys) in graphene. The Hamiltonian for a single valley ($\tau = \pm 1$ for $K/K'$) is:
\begin{equation}
    H = \hbar v_F \, \vec{\sigma} \cdot \vec{p},
\end{equation}
where:
\begin{itemize}
    \item $\hbar$ is the reduced Planck's constant,
    \item $v_F \approx 10^{6} \, \text{m/s}$ is the Fermi velocity,
    \item $\vec{\sigma} = (\sigma_x, \sigma_y)$ are the Pauli matrices (acting on the two sublattice pseudospin components A and B),
    \item $\vec{p} = -i\hbar (\partial_x, \partial_y)$ is the momentum operator.
\end{itemize}

This Hamiltonian describes massless Dirac fermions with linear dispersion:
\begin{equation}
    E = \pm \hbar v_F k,
\end{equation}
in infinite graphene.

\subsection{ Confinement in a Quantum Dot}

For a circular GQD of radius $R$, we assume the potential is zero inside ($r<R$) and infinite outside (modeling confinement via an \emph{infinite mass} boundary to prevent Klein tunneling). In polar coordinates $(r,\theta)$, the wavefunction is a two-component spinor:
\begin{equation}
    \Psi^\tau(r,\theta) = 
    \begin{pmatrix}
        \psi_A(r,\theta) \\
        \psi_B(r,\theta)
    \end{pmatrix}
    = e^{i m \theta}
    \begin{pmatrix}
        \chi_A(r) \\
        i e^{i\theta} \chi_B(r)
    \end{pmatrix},
\end{equation}
where $m = 0, \pm 1, \pm 2, \dots$ is the angular momentum quantum number (total angular momentum $j = m + 1/2$).  

Substituting into the Dirac equation $H\Psi = E\Psi$ yields coupled radial equations:
\begin{align}
    -\partial_r \chi_B + \frac{m+1}{r} \chi_B &= \frac{E}{\hbar v_F} \chi_A, \\
    \partial_r \chi_A + \frac{m}{r} \chi_A &= \frac{E}{\hbar v_F} \chi_B.
\end{align}

These can be decoupled into second-order equations resembling Bessel's differential equation.

\subsection{ Solutions and Energy Levels}

The radial solutions inside the dot are Bessel functions of the first kind:
\begin{equation}
    \chi_A(r) = J_m(k r), \quad \chi_B(r) = \pm J_{m+1}(k r),
\end{equation}
where $k = |E| / (\hbar v_F)$, and the sign depends on the energy sign (positive for conduction band, negative for valence band).

To derive the energy levels:

\begin{itemize}
    \item At the boundary $r=R$, the infinite mass condition ensures no probability current outflow. For the $K$ valley ($\tau = +1$), this gives:
    \begin{equation}
        \frac{\chi_B(R)}{\chi_A(R)} = -i e^{i \theta} 
        \quad \Rightarrow \quad 
        J_m(kR) = -J_{m+1}(kR).
    \end{equation}
    \item For the $K'$ valley ($\tau = -1$), it is:
    \begin{equation}
        J_m(kR) = J_{m+1}(kR).
    \end{equation}
\end{itemize}

The energy levels $E_{m,n}^\tau$ are then:
\begin{equation}
    E_{m,n}^\tau = \pm \frac{\hbar v_F}{R} \, \xi_{m,n}^\tau,
\end{equation}
where $\xi_{m,n}^\tau$ is the $n$-th root ($n = 1,2,3,\dots$) of the transcendental equation:
\begin{equation}
    J_m(\xi) = \mp J_{m+1}(\xi)
\end{equation}
(upper sign for $\tau = +1$, lower sign for $\tau = -1$). These roots can be solved numerically; for example, the lowest root for $m=0$ is approximately $\xi \approx 2.48$ for one valley.

The bandgap (energy separation between the highest valence and lowest conduction states) is approximately:
\begin{equation}
    E_g \approx \frac{2 \hbar v_F}{R},
\end{equation}
showing inverse dependence on size due to quantum confinement.

\subsection{ Extensions and Approximations}

\begin{itemize}
    \item \textbf{With Magnetic Field ($B$):} Include vector potential $\vec{A} = (B r/2) \hat{\theta}$. The radial equation becomes a confluent hypergeometric equation, leading to energy levels involving Laguerre polynomials:
    \begin{equation}
        E_{n,m} = \hbar v_F 
        \sqrt{\frac{2 e B}{\hbar c} 
        \left( n + |m| + \frac{1}{2} \right)},
    \end{equation}
    adjusted by boundary matching.
    \item \textbf{Tight-Binding Alternative:} For atomic-scale precision, use the nearest-neighbor tight-binding Hamiltonian on a hexagonal lattice:
    \begin{equation}
        H = -t \sum_{\langle i,j \rangle} 
        \left( c_i^\dagger c_j + \text{h.c.} \right),
    \end{equation}
    where $t \approx 2.8$~eV is the hopping parameter, and the sum is over nearest neighbors. For a finite GQD, construct the Hamiltonian matrix (size proportional to number of atoms $N \approx \pi R^2 / a^2$, $a=0.142$~nm lattice constant), and diagonalize to get eigenvalues. The bandgap scales as:
    \begin{equation}
        E_g \approx \frac{2t}{\sqrt{N}}.
    \end{equation}
\end{itemize}

This Dirac model captures the essential physics of GQDs' electronic properties. To compute specific levels, solve the Bessel root equation numerically (e.g., using Python's \texttt{scipy.special.jn\_zeros}, though adapted for the ratio). For example, for $m=0$, solve $J_0(\xi) + J_1(\xi) = 0$ iteratively starting from $\xi=2$.

\section{Comparison between GQDS and Traditional Quantum Dots like CDTE}

\begin{table}[h]
\caption{Physical \& Chemical Properties of GQDs and CdTe QDs}\label{tab1}%
\begin{tabular}{@{}lll@{}}
\toprule
Parameter & GQDs & CdTe QDs \\
\midrule
Toxicity & Non-toxic, biodegradable & Cd$^{2+}$ leaching risks \\
Bandgap Tuning & Size, edges, functionalization & Size-dependent only \\
Environmental Impact & Eco-friendly synthesis & Heavy-metal pollution \\
Cost & \$10--\$50/g (citric acid precursors)\cite{roushani2015graphene} & 
\$100--\$500/g (Cd/Te precursors)\cite{roushani2015graphene} \\
\bottomrule
\end{tabular}
\end{table}

\begin{table}[h]
\caption{Comparative Analysis of GQD and CdTe QD Properties}\label{tab2}%
\begin{tabular}{@{}lll@{}}
\toprule
Property & GQDs & CdTe QDs \\
\midrule
Composition & Carbon (sp$^{2}$), non-toxic & CdTe (toxic heavy metals) \\
Size Range & 1--10 nm (typically 2--5 nm)\cite{liu2015graphene} & 
1.2--6 nm (typically 2--5 nm)\cite{balkanloo2023graphene} \\
Bandgap Energy & 2.0--4.1 eV (size-dependent)\cite{liu2015graphene} & 
1.5--2.8 eV (size-dependent)\cite{balkanloo2023graphene} \\
Surface Groups & Abundant --OH, --COOH, --NH$_2$ & Requires ligand/polymer coating \\
Water Solubility & High (intrinsic) & Requires surface modification \\
Stability & High photostability & Prone to oxidation and photobleached \\
\bottomrule
\end{tabular}
\end{table}

\begin{table}[h]
\caption{Optical \& Electronic Properties of GQDs and CdTe QDs}\label{tab3}%
\begin{tabular}{@{}lll@{}}
\toprule
Property & GQDs & CdTe QDs \\
\midrule
PLQY & Up to 94\% (with doping) & 40--90\% \\
Emission Range & UV to NIR (tunable) & Visible to NIR (size-dependent) \\
Carrier Mobility & $>$24 cm$^{2}$/V$\cdot$s (ordered structures) & 1--10 cm$^{2}$/V$\cdot$s \\
Single-Photon Emission & Demonstrated at cryogenic temps & Limited by stability \\
\bottomrule
\end{tabular}
\end{table}
\begin{table}[h]
\caption{Applications of GQDs and CdTe QDs}\label{tab4}%
\begin{tabular}{@{}lll@{}}
\toprule
Application & GQDs & CdTe QDs \\
\midrule
Biosensing & Detect mycotoxins at 0.1 ppb & Limited by toxicity \\
Food Safety & Smart packaging for FCI storage & Rare due to safety concerns \\
Photovoltaics & Enhance light harvesting (PCE $>$ 12\%) & Established (PCE $>$ 15\%) \\
Quantum Computing & Spin qubits with long coherence times & Limited by nuclear spin interactions \\
\bottomrule
\end{tabular}
\end{table}

\newpage
\section{Conclusion}
GQDs represent a revolutionary advancement in biosensor technology, offering unprecedented advantages over traditional quantum dots due to their exceptional biocompatibility, superior photostability, and cost-effective synthesis from abundant carbon precursors. Their unique combination of quantum confinement effects, high surface-to-volume ratios, and tunable bandgaps (2.0--4.1 eV) enables detection limits in the femtomolar to picomolar range up to 1000$\times$ more sensitive than conventional biosensors while maintaining excellent selectivity and rapid response times (typically $<$15 seconds) \cite{mcardle2024new,mansuriya2020applications}.
The integration of GQDs with IoT infrastructure and quantum-optimized sensor networks demonstrates remarkable potential for real-world applications, from mycotoxin detection in food storage (achieving 0.158 ng/mL sensitivity for aflatoxin B1) to environmental monitoring systems capable of detecting heavy metals at nanomolar concentrations \cite{bera2010quantum}. As multifunctional platforms combining fluorescence imaging, electrochemical sensing, and therapeutic capabilities, GQDs are poised to transform personalized medicine, environmental monitoring, and food safety applications. Ongoing research is focusing on standardized synthesis protocols and regulatory frameworks to enable widespread clinical and commercial adoption \cite{singh2022graphene,rasheed2024graphene}.
Beyond these capabilities, GQDs demonstrate clear technical superiority with detection limits in the picomolar to nanomolar range across diverse analytes, rapid response times often below 15 minutes and sometimes as low as 3 seconds, and multiplexing potential for simultaneous detection of multiple targets. Practically, their cost-effective synthesis from abundant carbon precursors, non-toxic nature, and compatibility with existing analytical platforms and IoT systems make them highly feasible for scalable applications. Looking ahead, GQDs hold promise for quantum-enhanced sensing with improved signal-to-noise ratios, AI-integrated data analysis for pattern recognition, and autonomous, real-time environmental and health monitoring networks, underscoring their transformative potential across biomedical, environmental, and industrial domains.

\end{document}